\documentclass[a4paper,11pt]{article}
\pdfoutput=1 
\usepackage{jcappub} 
\usepackage[T1]{fontenc} 
\usepackage{physics}
\usepackage{wrapfig}
\usepackage{xcolor}
\usepackage{float}
\restylefloat{table}
\usepackage{changepage}
\usepackage{booktabs}
\usepackage{subcaption}
\usepackage{graphicx}
\usepackage{csquotes}
\usepackage[normalem]{ulem}
\newcommand{\pvec}[1]{\vec{#1}\mkern2mu\vphantom{#1}}

\title{Effects of primordial fluctuations on relic neutrino simulations}

\author[a,1]{Fabian Zimmer,\note{Corresponding author.}}
\author[a]{Guillermo Franco Abellán,}
\author[a,b]{and Shin'ichiro Ando}

\affiliation[a]{GRAPPA Institute, University of Amsterdam, Science Park 904, 1098 XH Amsterdam, The Netherlands}
\affiliation[b]{Kavli Institute for the Physics and Mathematics of the Universe, University of Tokyo, Chiba 277-8583, Japan}

\emailAdd{f.zimmer@uva.nl}
\emailAdd{g.francoabellan@uva.nl}
\emailAdd{s.ando@uva.nl}

\abstract{After decoupling, relic neutrinos traverse the evolving gravitational imhomogeneities along their trajectories. Once they turn non-relativistic, this results in a significant amplification of the anisotropies in the cosmic neutrino background (C$\nu$B). Past studies have reconstructed the phase-space distribution of relic neutrinos from the local distribution of matter (accounting for the Milky Way halo and the surrounding large-scale structures), but have neglected the C$\nu$B anisotropies in the initial conditions of neutrino trajectories. Using our previously developed N-1-body simulation framework, we show that including these primordial fluctuations in the initial conditions can be important, as it produces similar effects on the abundance and anisotropies of the C$\nu$B as the inclusion of large-scale structures beyond the Milky Way halo. Interpretability of data from future C$\nu$B observatories like PTOLEMY therefore depends on correctly modelling these effects.

\vspace*{5pt} \noindent \textbf{\texttt{GitHub}}: Our \texttt{jax}-accelerated simulation code can be found \href{https://github.com/Fabian-Zimmer/neutrino_clustering.git}{here}.
}


\begin{document}
\maketitle
\flushbottom

\section{Introduction}
\label{sec:intro}
A fundamental prediction of the concordance cosmological model is the presence of a cosmic background of relic neutrinos (C$\nu$B), formed when they decoupled from the primordial plasma only one second after the Big Bang. While there is strong indirect evidence of the C$\nu$B around the Big Bang Nucleosynthesis (BBN) \cite{Pisanti:2020efz,Fields:2019pfx} and the Cosmic Microwave Background (CMB)~\cite{Follin:2015hya,Baumann:2019keh} eras, little is known about their properties. In particular, their absolute mass scale and mass ordering are still unknown, but are projected to be measured in the coming years (see e.g.~\cite{Gerbino:2022nvz}). Their background phase-space distribution is assumed to be of the Fermi-Dirac type~\cite{Kolb:1990vq,Lesgourgues:2013sjj,Baumann:2022mni} as per theoretical expectations, although there exist only weak cosmological bounds on relic neutrino statistics~\cite{deSalas:2018idd,Alvey:2021sji}.

The direct detection of the C$\nu$B in a laboratory setting also remains a huge challenge. Future experiments such as PTOLEMY~\cite{PTOLEMY:2019hkd} aim to detect the C$\nu$B via neutrino capture on $\beta$-decaying nuclei, in particular tritium. The event rate in these experiments is proportional to the local number density of relic neutrinos, which is expected to be larger than the cosmological average due to gravitational effects in the vicinity of the Earth. If the tritium targets are polarized, this would possibly allow the measurement of C$\nu$B anisotropies \cite{Lisanti:2014pqa}, which are also expected to be influenced by the local gravitational environment. Hence, to understand the prospects of these future experiments, previous works have investigated gravitational clustering of relic neutrinos via linear perturbation theory~\cite{Singh:2002de,Alvey:2021xmq,Holm:2023rml,Holm:2024zpr} or N-body or N-1-body simulation frameworks~\cite{Ringwald:2004np,deSalas:2017wtt,Zhang:2017ljh,Mertsch:2019qjv,Elbers:2023mdr,Zimmer:2023jbb}.\ 

However, all these works have so far assumed  a perfectly homogenous and isotropic Fermi-Dirac distribution for neutrinos right before the onset of non-linear structure formation. In practice, we expect there to be some non-negligible fluctuations in the phase-space distribution which are induced by the primordial fluctuations in the matter fields. Using linear theory, it was shown in~\cite{Tully:2021key} that such primordial fluctuations lead to anisotropies in the C$\nu$B temperature that can grow up to ${\sim}10\%$ of the mean C$\nu$B temperature of $T_{\nu,0} \sim 1.95$~K.\  

\begin{figure}[t!]
    \centering
    \includegraphics[width=0.8\textwidth]{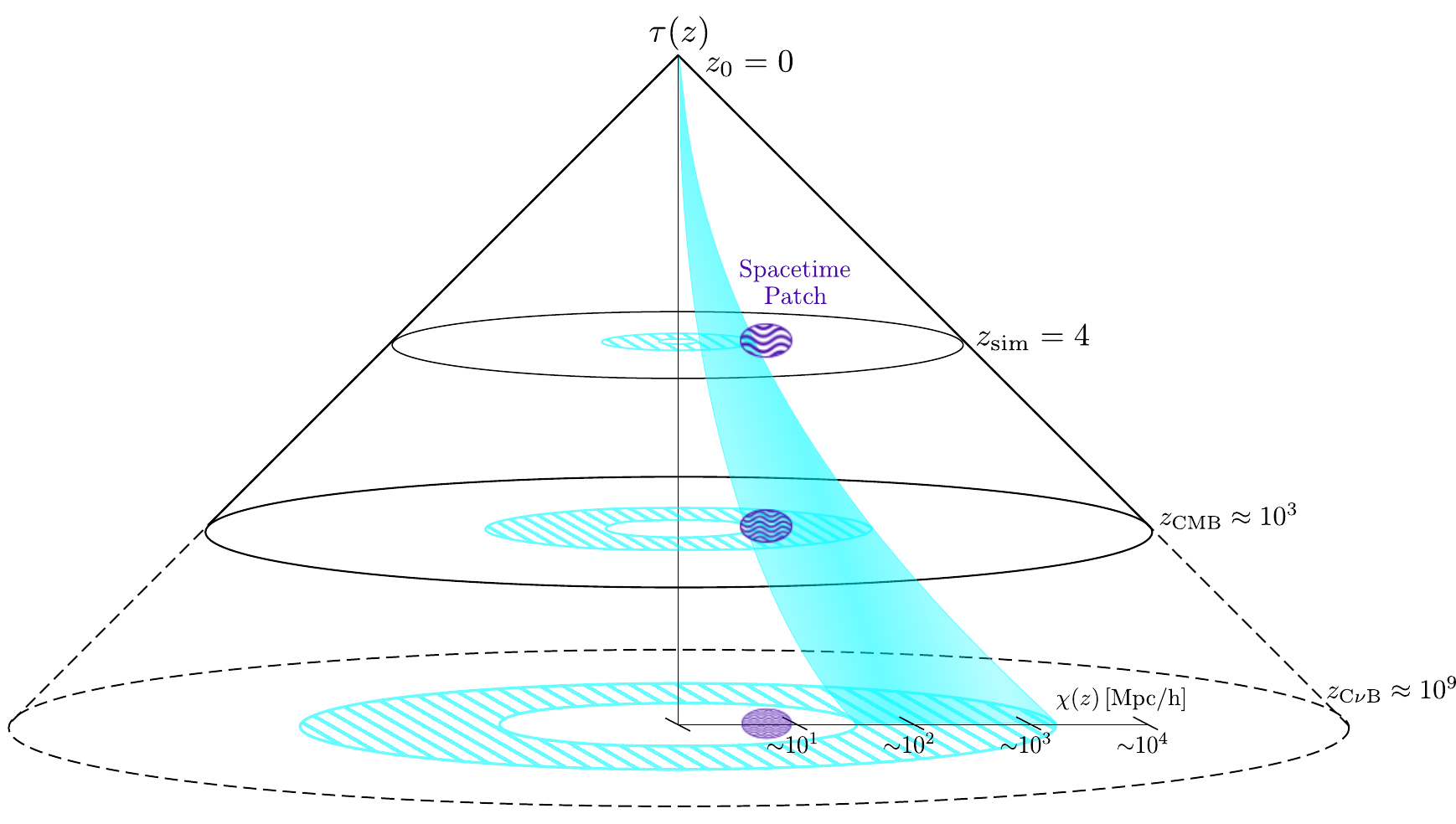}
    \caption{Schematic diagram of our past lightcone, extrapolated until the time of neutrino decoupling. The cyan band roughly represents $0.05 \, \mathrm{eV}$ neutrinos with comoving momenta in the range of about $[0.01,10]$ times the present C$\nu$B temperature. Note that the spacings and areas are not to scale; the figure is merely aimed to be a helpful visualization.}
    \label{fig:neutrino_cone}
\end{figure}

The C$\nu$B anisotropies exhibit a very rich phenomenology that presents several important differences compared to those of the CMB. To illustrate this, we show a schematic diagram of our past lightcone including neutrinos in Figure~\ref{fig:neutrino_cone}. The horizontal direction represents comoving distance and the vertical direction conformal time.  
In addition to our spacetime point at the tip of the cone (at $z_0 = 0$), we show three other constant-time hypersurfaces at $z_\mathrm{sim} = 4$ (the time boundary of our simulations, as explained later), $z_\mathrm{CMB} \approx 10^3$ (roughly the time of photon decoupling) and $z_{\mathrm{C}\nu\mathrm{B}} \approx 10^9$ (roughly the time of neutrino decoupling).
The lightlike worldline is depicted with the solid black lines at a $45^\circ$ angle, and we extended this lightcone beyond the CMB for illustrative purposes with the dashed black lines. The cyan band approximately represents $0.05 \, \mathrm{eV}$ neutrinos with comoving momenta in the range of $0.01$ to $10$ times the present C$\nu$B temperature. At neutrino decoupling they are ultra-relativistic, with an angle of almost $45^\circ$, then slow down due to the cosmic expansion, which corresponds to a progressively increasing change of the angle as they become more and more non-relativistic at low redshifts. The hatched cyan shell elements represent the domains of space which relic neutrinos of this momentum band traverse. The fact that the C$\nu$B originates from a last-scattering surface closer to us than the CMB~\cite{Bisnovatyi:1083,Bisnovatyi:1084,Dodelson:2009ze}, where the distance is determined by the neutrino momentum, is clearly visible in this diagram: the CMB photons travelled $\sim 10^4~\mathrm{Mpc}$, whereas the majority of relic neutrinos travelled $\sim 10^3~\mathrm{Mpc}$ (in comoving units). This means that neutrinos with different masses, or equivalently, neutrinos of the same mass but with different momenta pass through the same ``spacetime patch'' (illustrated in purple) at different times.\footnote{An interesting consequence is that the three neutrino mass eigenstates get lensed differently, which theoretically can be used to probe the lensing source at different times~\cite{Lin:2019lko}, allowing for a kind of neutrino tomography of LSS.} As a result, the temperature anisotropies differ between neutrino momenta (and masses), as they are affected by different evolving gravitational potentials at distinct times.\footnote{At most we can hope to eventually probe the momentum averaged C$\nu$B anisotropies as computed in~\cite{Michney:2006mk,Hannestad:2009xu,Tully:2021key} and studied specifically in relation to the PTOLEMY experiment~\cite{Betts:2013uya,Long:2014zva,PTOLEMY:2019hkd} in~\cite{Lisanti:2014pqa,Tully:2022erg}.} The framework developed in~\cite{Tully:2021key} computes these C$\nu$B anisotropies.

In our previous work~\cite{Zimmer:2023jbb}, we constructed a numerical simulation framework based on the approach of~\cite{Mertsch:2019qjv}, to compute the local C$\nu$B density as well as its anisotropies. This was done by initializing relic neutrinos at Earth and reconstructing their trajectories through the local gravitational environment by running the simulations backwards to a time, where the neutrino phase-space was assumed to be Fermi-Dirac. Liouville's theorem maps this phase-space to the (unknown) one today, allowing the moments of the distribution like e.g. the number density to be computed. Even though our simulation space in that study only contained dark matter (DM) halos resembling the Milky Way (MW), without the large-scale structure (LLS), we uncovered effects not seen with previous analytical treatments: the neutrinos could be ``diffused'' by DM particles along their trajectories, producing anti-correlations between the neutrino number densities and the projected DM content along the line-of-sight. We speculated that the inclusion of LSS might enhance this effect, depending on how much DM structure a neutrino of a given momentum traverses.

In this work, we describe an extension to this framework to partially account for the effects of LSS $-$ motivated by the previous discussion and the most recent work of C$\nu$B clustering~\cite{Elbers:2023mdr}. In that latest study they employed both full N-body and N-1-body simulation techniques to study relic neutrino clustering, fully including LSS inside their $1 \, \mathrm{Gpc}^3$ simulation volume. Amongst other things, they showed that distant DM structures can be anti-correlated with local neutrino fluctuations in the same direction on the sky, with the effect being more evident for smaller neutrino masses which traverse greater distances. The net effect of including LSS appears to almost always\footnote{For the smallest neutrino mass of $0.01$ eV they find a very small deficit instead.} contribute positively to the number density, with the fractional contribution to the total density being dependent on the neutrino mass. To make precise predictions about the local C$\nu$B properties it thus seems crucial to include the LSS of the Universe. We show that when modifying the boundary conditions of our simulation by including anisotropic phase-space distributions from linear primordial fluctuations, we recover some of the effects that one would obtain by fully modeling the LSS within the N-body simulation domain. 

The remainder of this work is structured to reflect the order of execution to arrive at the results. The methodology (section~\ref{sec:methods}) is divided into three parts: how we reconstruct the neutrino trajectories and compute the number densities (section~\ref{sec:CNB_sims}), how we compute the C$\nu$B anisotropies (section~\ref{sec:C_l}), and how we build the new boundary conditions for the neutrino trajectories based on these anisotropies (section~\ref{sec:boundary_conds}). The results are then presented in section~\ref{sec:results} and we disuss and conclude in section~\ref{sec:discussion_and_conclusions}.

\section{Methods}
\label{sec:methods}

In this section we give an overview of our simulation framework, for which more details can be found in~\cite{Zimmer:2023jbb}. We then show how to compute the C$\nu$B temperature anisotropies with the framework developed in~\cite{Tully:2021key} and describe how to change the boundary conditions for our simulations accordingly.

\subsection{C\texorpdfstring{$\nu$}{nu}B simulations}
\label{sec:CNB_sims}

Our simulations reconstruct relic neutrino trajectories, which are governed by Hamilton's equations of motion
\begin{equation} \label{eq:EOMs}
\vec{q}_j = a m_\nu \vec{x}_j', \quad \pvec{q}_j'= -a m_\nu \frac{\partial \Phi(\vec{x}, t)}{\partial \vec{x}_j},
\end{equation}
where $\vec{q}_j$ is the comoving momentum of the $j$-th neutrino and $\vec{x}_j$ its position. The prime notation of these quantities denotes the derivative with respect to conformal time. The scale factor $a$ accounts for the cosmic expansion, and any effect due to non-linear structure formation enters via the gravitational potentials $\Phi$ at position $\vec{x}$ and time $t$. These potentials are formed by DM halos selected from the {\sc TangoSIDM} project \cite{Correa:2022dey,Correa:2024vgl} and have masses consistent with the estimated Milky-Way mass using the Gaia data \cite{Karukes:2019jwa,2019A&A...621A..56P,2019ApJ...875..159E,2018A&A...619A.103F}, whose values are within the range of $M_{200} = (0.6\mbox{--}2.0) \times 10^{12} \, \mathrm{M_\odot}$. Here, we use 26 such halos whose merger and accretion history is represented by how the DM particles comprising the halo are distributed at various redshifts. The DM distribution at a redshift is referred to as a snapshot, and we use 25 of them between redshifts $0$ and $4$. We run the simulation backwards in time, which implies that the neutrinos are initiated in a cell with a distance from the halo center akin to the distance of Earth to the MW Galactic center. They are then aimed in different directions corresponding to the pixels of a healpy allsky map with a resolution of $\mathrm{Nside}=8$ (768 pixels). There are multiple cells fitting this distance criterion, but we showed in~\cite{Zimmer:2023jbb} that choosing a different starting cell does not change the overall results much. We upgraded our previous code using the {\tt jax} programming paradigm~\cite{jax2018github} to accelerate the simulations, as well as to solve the equations of motion with the {\tt jax}-compatible {\tt diffrax} library. This allowed us to both increase precision and decrease computation time. We found that simulating 1000 neutrinos for each direction in the sky with this methodology was sufficient. The momenta are spaced logarithmically in the range $[0.01~T_{\nu,0},~ 400~T_{\nu,0}]$. We used a 128 CPU core cluster node with 224 GB memory to run the simulations, resulting in a runtime of $\mathcal{O}(1)$ minutes for a single halo.

Such simulations allow us to find the final positions $\vec{x}_j(z_{\rm sim})$ and momenta $\vec{q}_j(z_{\rm sim})$ of the neutrinos at the last redshift of our simulations, $z_{\rm sim}$. Liouville's theorem states that the phase-space density is conserved along particle trajectories, so this allows us to compute the present values of the phase-space as 
\begin{equation} \label{eq:Liouville}
f(\vec{x}_\oplus, \vec{q}_j(0), 0) = f_{\rm sim} (\vec{x}_j(z_{\rm sim}), \vec{q}_j(z_{\rm sim}), z_{\rm sim}),
\end{equation}
where $\vec{x}_\oplus$ is the starting coordinate of all neutrinos (i.e. the position of Earth) and $f_{\rm sim}$ represents the initial phase-space distribution of the neutrinos at $z_{\rm sim}$. The local number density per degree of freedom is then computed as
\begin{equation} \label{eq:n_nu}
n_\nu = \int \frac{d^3 \vec{q}_0}{(2 \pi)^3} \, f(\vec{x}_\oplus, \vec{q}_0, 0),
\end{equation}
where the integral should be understood as a sum over the momenta $\vec{q}_j(0)$ of all simulated neutrinos.

\subsection{Angular power spectrum}
\label{sec:C_l}

In our previous study \cite{Zimmer:2023jbb} we had assumed a Fermi-Dirac distribution to describe the state of the C$\nu$B at the last redshift of our simulations, i.e. $f_{\rm sim} = f_{\rm FD}$ in eq.~(\ref{eq:Liouville}). We now go over the equations from cosmological perturbation theory necessary to compute the anisotropies in the C$\nu$B temperature, which are induced by the evolving gravitational inhomogeneities along the trajectories of the relic neutrinos. This will ultimately give us a perturbed phase-space distribution, $f_\Delta$, which serves as a proxy for the effects from LSS. For this, we adapted the line-of-sight method from~\cite{Tully:2021key}. We refer the reader to that study for details omitted in this section.

We start by introducing a perturbation $\Psi$ in the Fermi-Dirac distribution for neutrinos to get the perturbed distribution
\begin{equation} \label{eq:f_FD_Psi}
    f(\vec{x}, q,\hat{n}, \tau) = f_{\rm FD}(q) (1 + \Psi(\vec{x}, q,\hat{n}, \tau)) = \frac{1}{e^{\frac{q}{T_{\nu,0}}}+1} (1 + \Psi(\vec{x}, q,\hat{n}, \tau)),
\end{equation}
where the vector $\hat{n}$ represents the direction of comoving momenta. If we now write the left-hand side in terms of a fractional temperature perturbation $\Delta$,
\begin{equation} \label{eq:f_perturbed}
    f_\Delta(q)=\left(e^{\frac{q}{T_{\nu,0}(1+\Delta)}}+1\right)^{-1} 
\end{equation}
and expand this to linear order, we can express $\Delta$ in terms of $\Psi$ as 
\begin{equation} \label{eq:Delta}
    \Delta=-\left(\frac{d \ln f_{\rm FD}}{d \ln q}\right)^{-1} \Psi.
\end{equation}
Note that the neutrino temperature perturbation $\Delta$ depends not only on space and time, but also on momenta, meaning that the perturbed phase-space $f_\Delta$ acquires spectral distortions. From now on, we will only display the momentum dependence for simplicity. To eliminate the dependence on direction it is customary to switch to Fourier space\footnote{From now on, we will use tilde to denote quantities in Fourier space.} and expand $\tilde{\Psi}$ in Legendre multipoles; the resulting neutrino perturbations per multipole, $\tilde{\Psi}_\ell$, obey a Boltzmann hierarchy of equations that can be solved numerically with Boltzmann solvers like {\tt CLASS}~\cite{Blas:2011rf}. Then, from $\tilde{\Psi}_\ell$ one can obtain the associated temperature perturbations $\tilde{\Delta}_\ell$ via eq.~(\ref{eq:Delta}). These in turn can be used to get the angular power spectrum with
\begin{equation} \label{eq:Cls_los}
    C_\ell (q) = 4 \pi T_{\nu,0}^2 A_s \int d \ln k \Tilde{\Delta}_\ell^2(q) ,
\end{equation}
where $A_s$ is the primordial power spectrum amplitude.\footnote{Eq.~(\ref{eq:Cls_los}) assumes a spectral index of $n_s=1$. We refer the reader to~\cite{Tully:2021key} and their public code available at \url{https://github.com/gemyxzhang/cnb-anisotropies} for more details on other cosmological parameters.}

The so-called line-of-sight method allows for a quick calculation of $\Tilde{\Delta}_\ell(q)$ up to a very large $\ell_{\rm max}$ without having to solve a Boltzmann hierarchy of $(\ell_{\rm max}+1)$ equations. More precisely, this method expresses $\Tilde{\Delta}_\ell(q)$ as an integral from neutrino decoupling ($\tau_{\mathrm{dec}}$) to the present day ($\tau_0)$, namely
\begin{equation} \label{eq:Delta_l}
    \Tilde{\Delta}_\ell(q) = \frac{1}{2} \phi_{\rm dec} j_\ell(k \chi(z_{\rm dec})) + \int_{\tau_{\rm dec}}^{\tau_0} d \tau \left\{ 2 \left(\frac{a m_\nu}{q}\right)^2 (aH) \phi + \left[2 + \left(\frac{a m_\nu}{q}\right)^2\right] \partial_\tau \phi \right\} j_\ell(k \chi(z_\tau)).
\end{equation}
In this equation, we see the spherical Bessel functions of the first kind $j_\ell$, the neutrino mass $m_\nu$ and the Hubble parameter $H$. 
The second term represents the amplification of anisotropies due to the evolving gravitational potential $\phi$ (in the conformal Newtonian gauge) encountered by neutrinos during their non-relativistic phase, while the first term with $\phi_\mathrm{dec}$ captures the anisotropies at neutrino decoupling.
Finally, the Bessel functions depend on the comoving distance traveled by massive neutrinos from an initial redshift $z_i$ to a final redshift $z_f$ (with $z_i > z_f$) as given by
\begin{equation} \label{eq:Chi}
    \chi(z_i,q) = \int_{z_f}^{z_i} \frac{dz}{H(z)} \frac{q}{\sqrt{q^2 + m_\nu^2/(1+z)^2}}.
\end{equation}
Note the dependence of $\chi(z_i,q)$ on the ratio $m_\nu/q$, as we illustrated in Figure~\ref{fig:neutrino_cone}. The method also allows the momentum-averaged angular power spectrum to be expressed as
\begin{equation} \label{eq:Cls_los_qavg}
    C_\ell = 4 \pi T_{\nu,0}^2 A_s \int d \ln k \left( \int dq \, \Tilde{\Delta}_\ell(q) \, \frac{2}{3\zeta(3)T^3_{0}} \, \frac{q^2}{e^{\frac{q}{T_{0}}} + 1} \right)^2 .
\end{equation}

\noindent Let us also note that the line-of-sight method is strictly only valid for neutrino masses of  $m_\nu \lesssim 0.1 \  \mathrm{eV}$ and multipoles $\ell \lesssim 30$. For larger neutrino masses and multipoles, the predictions from the line-of-sight method can deviate significantly from the full Boltzmann hierarchy calculation, as shown in \cite{Tully:2021key}. For this reason, our main results concerning the C$\nu$B anisotropies (which rely on the line-of-sight formalism) are presented only for multipoles up to $\ell \sim 20$ and neutrino masses up to $0.1 \  \mathrm{eV}$\footnote{More generally, linear theory breaks down at $z=0$ for neutrino masses above $0.1 \  \mathrm{eV}$, as we will show explicitely in Figure~\ref{fig:Delta_vs_z}.}. 

\subsection{Boundary conditions}
\label{sec:boundary_conds}

\begin{figure}[t!]
    \centering
    \includegraphics[width=0.8\textwidth]{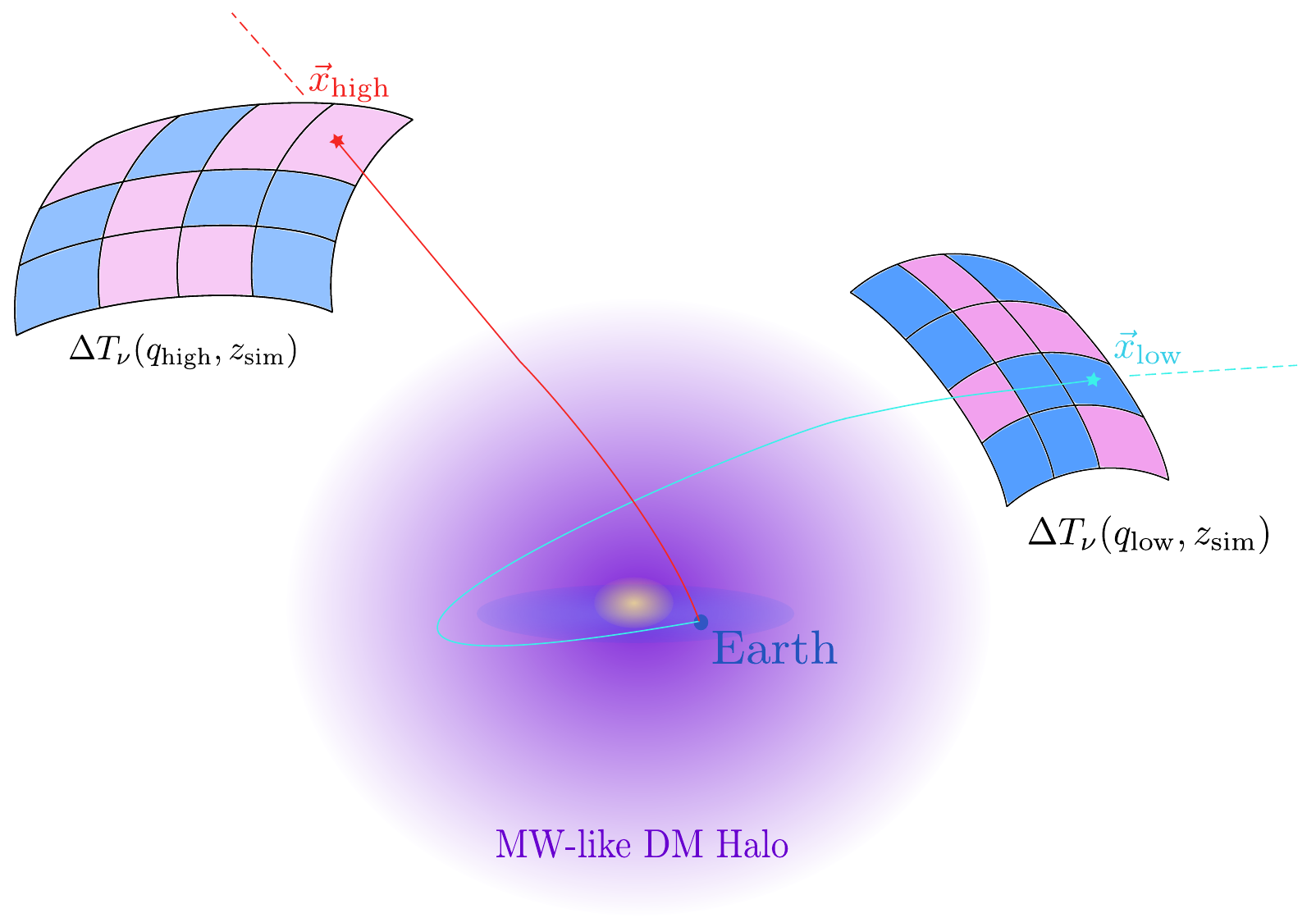}
    \caption{Schematic illustration of the temperature fluctuation surfaces (colored grids) at the last redshift of our simulations of $z_\mathrm{sim} = 4$ for fast- ($q_\mathrm{high}$) and slow-moving ($q_\mathrm{low}$) neutrinos (see text for more details concerning the connection between the surfaces and neutrino trajectories). The Milky Way disc and bulge, as depicted by the blue and yellow oval shapes, are for visualization purposes only, since our simulations do not include baryonic components.}
    \label{fig:Cl_surfaces}
\end{figure}
We can use the expressions for the angular power spectrum of the previous section to construct more accurate boundary conditions for our simulations. Technically, the spatial boundary varies for each neutrino: the ones initialized with a high momentum ($q_\mathrm{high}$) end up further away from Earth than those starting with a low momentum ($q_\mathrm{low}$). We therefore have to find the correct value of $\Delta$ for each neutrino. For the following explanation, it is again helpful to turn to an illustration. 
Figure~\ref{fig:Cl_surfaces} schematically shows patches of two different temperature fluctuation surfaces depicted by the curved pink and blue colored grids. These illustrate possible temperature fluctuation patterns arising from the angular power spectrum in eq.~(\ref{eq:Cls_los}) for the two different neutrino momenta. The red and cyan curves exemplify trajectories of neutrinos with $q_\mathrm{high}$ and $q_\mathrm{low}$, respectively. The last positions of these neutrinos (i.e. at $z_{\rm sim}$) are marked with $\vec{x}_\mathrm{high}$ and $\vec{x}_\mathrm{low}$, where they have momenta $q_{\rm high}^{z_{\rm sim}}$ and $q_{\rm low}^{z_{\rm sim}}$, respectively.

\begin{figure}[t!]
    \centering
    \includegraphics[width=.7\textwidth]{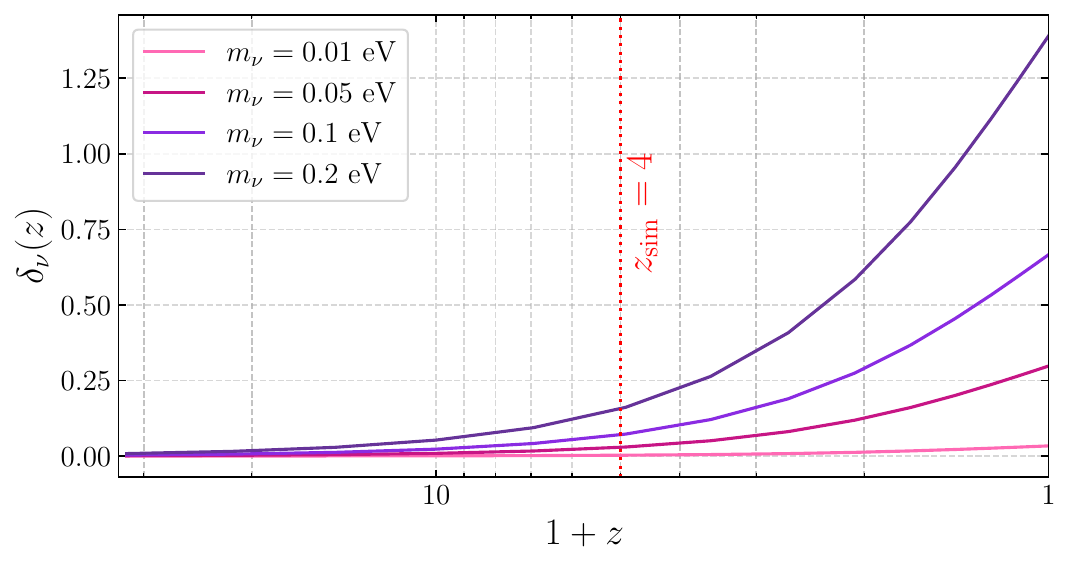}
    \caption{Approximate neutrino energy density contrast as a function of redshift for different neutrino masses of $0.01$, $0.05$, $0.1$ and $0.2$ eV (see text for details on the calculation of this quantity). The last redshift of our simulation of $z_{\rm sim}=4$ is marked with the red dotted line. The overdensities around that time can already reach ${\sim 20\%}$, highlighting the neccessity to account for these deviations from an isotropic and homogeneous phase-space distribution.}
    \label{fig:Delta_vs_z}
\end{figure}
Let us now rewind time to neutrino decoupling and follow two specific ``newly-released'' neutrinos with momenta $q^{\mathrm{C}\nu\mathrm{B}}_\mathrm{high}$ and $q^{\mathrm{C}\nu\mathrm{B}}_\mathrm{low}$, which will end up at these positions. Due to the influence of the primordial gravitational potential $\phi$, these momenta will evolve until reaching values $q_{\rm high}^{z_{\rm sim}}$ and $q_{\rm low}^{z_{\rm sim}}$, which characterize a linearly perturbed phase-space according to the formalism described in section~\ref{sec:C_l}. So to get the correct value of $\Delta$ for each neutrino at their boundary, we need to: i) plug their corresponding momentum at $z_{\rm sim}$ into eq.~(\ref{eq:Cls_los}), ii) generate the corresponding all-sky temperature anisotropy maps, and iii) select the pixel corresponding to their last position. Then, eq.~(\ref{eq:f_perturbed}) yields the corresponding phase-space value, and we just set $f_{\rm sim} = f_{\Delta}$ to compute the number density with eq.~(\ref{eq:Liouville}) and eq.~(\ref{eq:n_nu}). In this way we account for the effects of a random linear LSS configuration from neutrino decoupling at $z_{\mathrm{C}\nu\mathrm{B}} \approx 10^9$ until the last redshift of our time-reversed simulations at $z_{\rm sim}=4$. Notice that to get the approriate temperature fluctuation skymaps at $z_{\rm sim}$ we need to cut off the time integrals in eq.~(\ref{eq:Delta_l}). The computation of $\Tilde{\Delta}_\ell(q)$ via this equation is characterized by two independent terms: a term proportional to the gravitational potential at neutrino decoupling ($\phi_\mathrm{dec}$) and a term with an integral over time. The first term also depends on the comoving distance through the Bessel function, and hence also involves an integral over time.
By cutting off these time integrals at a specific time prior to today, $\tau_\mathrm{cut}$ (or $z_\mathrm{cut}$), we effectively stop the evolution of the primordial temperature anisotropies at that time. An extra multiplicative factor of $(1+z_\mathrm{cut})^2$ is also needed in eq.~(\ref{eq:Cls_los}) to obtain the correct value of the monopole temperature at $z_\mathrm{cut}$, since $T \propto a^{-1}$. This procedure gives us the angular power spectrum at that higher redshift, $C_\ell(z_\mathrm{cut})$, from which we can produce
temperature anisotropy skymaps\footnote{For simplicity, we use the same pseudo-random realisations of the anisotropy skymaps generated from these $C_\ell$ across all our momentum bins.} and ultimately obtain the perturbed phase-space values, $f_\Delta(z_{\rm cut})$.

To highlight the neccessity to include these more realistic boundary conditions for our simulations, we show the evolution of the neutrino energy density contrast $\delta_\nu$ for $0.01$, $0.05$, $0.1$ and $0.2$ eV neutrinos in figure~\ref{fig:Delta_vs_z}. 
To get an approximate estimate of $\delta_\nu(z)$, we generated skymaps\footnote{We used the {\tt healpy}~\cite{Gorski:2004by} library throughout this work.} of the temperature perturbations at different redshifts, as realised by eq.~(\ref{eq:Cls_los_qavg}) with the $z_{\rm cut}$-technique described above.
Then, to obtain $\delta T_\nu$ we took the mean of the 2 extremes around the monopole temperature (one positive and one negative), and converted this into a density contrast using $\delta_\nu (z) \sim 3\frac{\delta T_\nu}{T_\nu}(z)$, which holds for non-relativistic neutrinos.
From figure~\ref{fig:Delta_vs_z}, we see that the assumption of a Fermi-Dirac distribution at our simulation boundary redshift ($z_\mathrm{sim} = 4$) is not appropriate, as the density contrasts can already reach up to ${\sim} 20\%$ for certain neutrino masses. We will show below that this has consequences for the number densities as obtained with our simulation framework, and that by accounting for these initial density perturbations, we can partly mimic the effects one would obtain when fully including LSS (i.e. structures beyond the MW Galaxy) inside the simulation domain.

\section{Results}
\label{sec:results}

We now show the changes to the local number density and anistropies of the C$\nu$B when using a perturbed phase-space distribution instead of the isotropic Fermi-Dirac distribution as the boundary condition of our simulation. The results representing the former case are obtained when setting $f_{\rm sim}=f_\Delta$ (i.e. specific realizations of $f_\Delta$) in eq.~(\ref{eq:Liouville}), whereas setting $f_{\rm sim}=f_{\rm FD}$ represents the latter case.
\begin{figure}[t!]
    \centering
    \begin{subfigure}[b]{.48\textwidth}
        \centering
        \includegraphics[width=\linewidth, height=\linewidth]{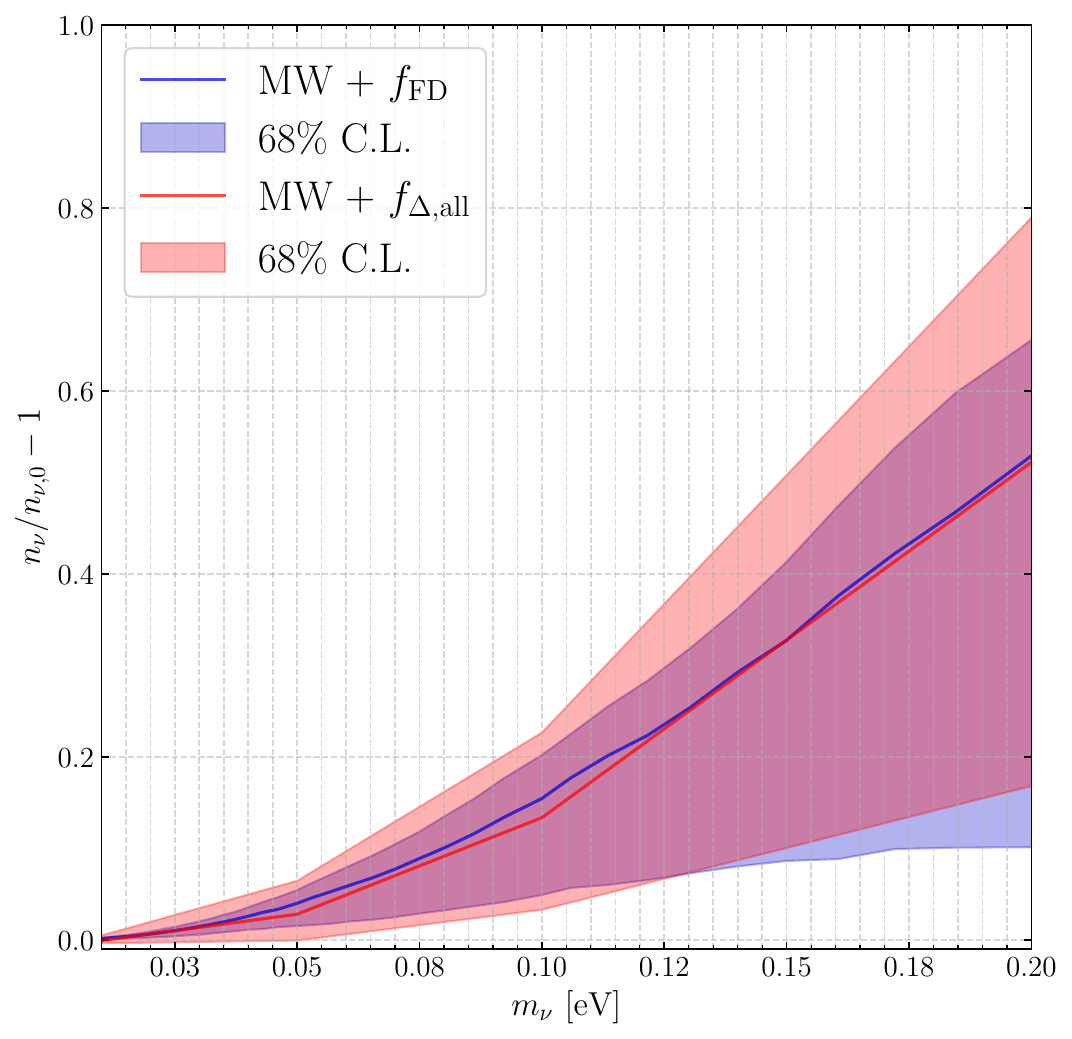}
    \end{subfigure}%
    \begin{subfigure}[b]{.48\textwidth}
        \begin{subfigure}[b]{\linewidth}
            \centering
            \includegraphics[width=\linewidth, height=.5\linewidth]{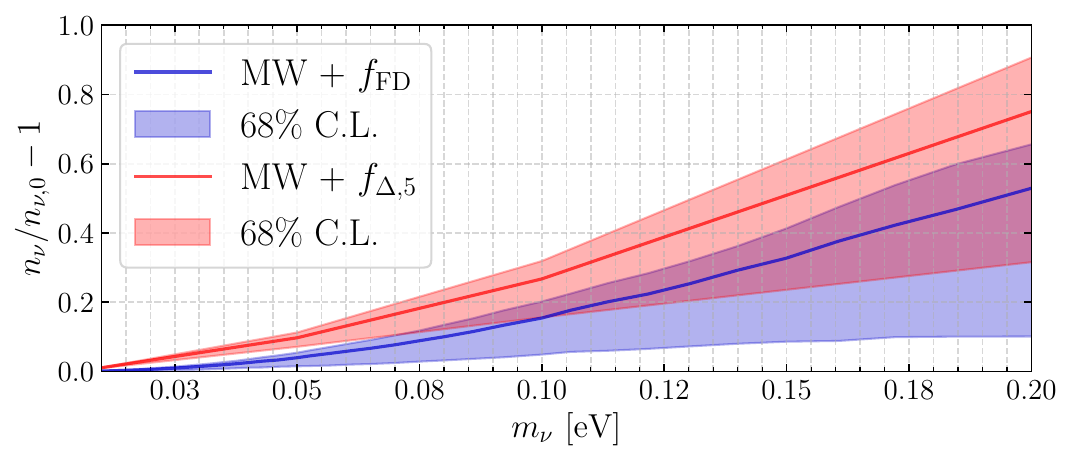}
        \end{subfigure}
        \vspace{-0.05cm}  
        \begin{subfigure}[b]{\linewidth}
            \centering
            \includegraphics[width=\linewidth, height=.5\linewidth]{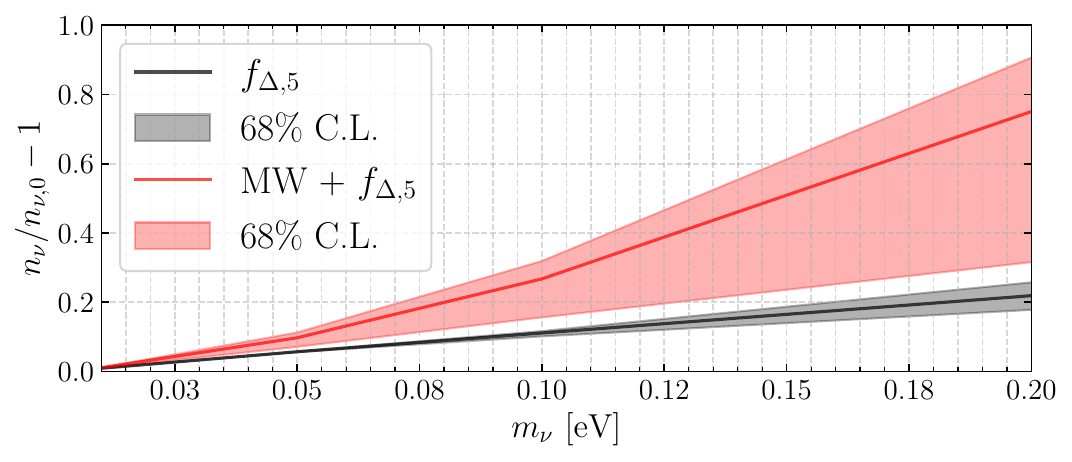}
        \end{subfigure}
    \end{subfigure}
    \caption{Neutrino clustering factors as a function of neutrino mass. \textbf{Left panel:} Using the Fermi-Dirac ($f_{\mathrm{FD}}$) boundary condition yields the blue band, where the dispersion is due to the different MW-like DM halos. Using 10 different perturbed phase-spaces ($f_{\Delta,{\rm all}}$) as boundary conditions for each DM halo in our sample gives the total uncertainty band in red. \textbf{Top right panel:} Uncertainty bands representing a specific LSS configuration ($f_{\Delta,5}$) that tends to increase the local number densities. \textbf{Bottom right panel:} The contribution solely due to the effects of this specific $f_{\Delta,5}$ is represented by the black band, and the red band is again shown for comparison.}
    \label{fig:eta_bands}
\end{figure}

\subsection{Local number density}

Figure~\ref{fig:eta_bands} displays the C$\nu$B number densities normalized to the cosmological average value (i.e. ${\sim}56 \, \mathrm{cm}^{-3}$ per mass eigenstate and per helicity) for a range of neutrino masses. The solid lines show the median and the shaded areas represent 68\% coverage bands. In blue we show the simulations using $f_\mathrm{FD}$ as the boundary condition and the uncertainty is due to the different DM halos. The red band in the left panel represents the combined uncertainty of using 10 realizations of $f_\Delta$ for each halo (denoted by $f_{\Delta,{\rm all}}$). 
There is no net increase or decrease for the local number density independent across all realisation of LSS in this way, as indicated by the similarity of both medians and uncertainty bands.
For the right panels we illustrate the effects of a specific LSS configuration, $f_{\Delta,5}$, which increases the number density on average for all neutrino masses, the effect being actually larger for smaller neutrino masses (top right panel), and where the effect of LSS alone, obtained simply by subtracting the blue band from the red, contributes only positively (shown as the black band in the bottom right panel). Our results for this specific case seem to agree with Figure 3 in \cite{Elbers:2023mdr}, where the authors show the individual contribution of LSS included in their simulations. For certain LSS configurations we recover the small deficits of order $\mathcal{O}(10^{-3})$ for neutrino masses of $\lesssim 0.05$ eV found in \cite{Elbers:2023mdr}. The change to more realistic boundary conditions thus has meaningful consequences and should be included for a N-1-body simulation framework such as ours.

\subsection{Anisotropies}

\begin{figure}[t!]
    \centering
    \includegraphics[width=\textwidth]{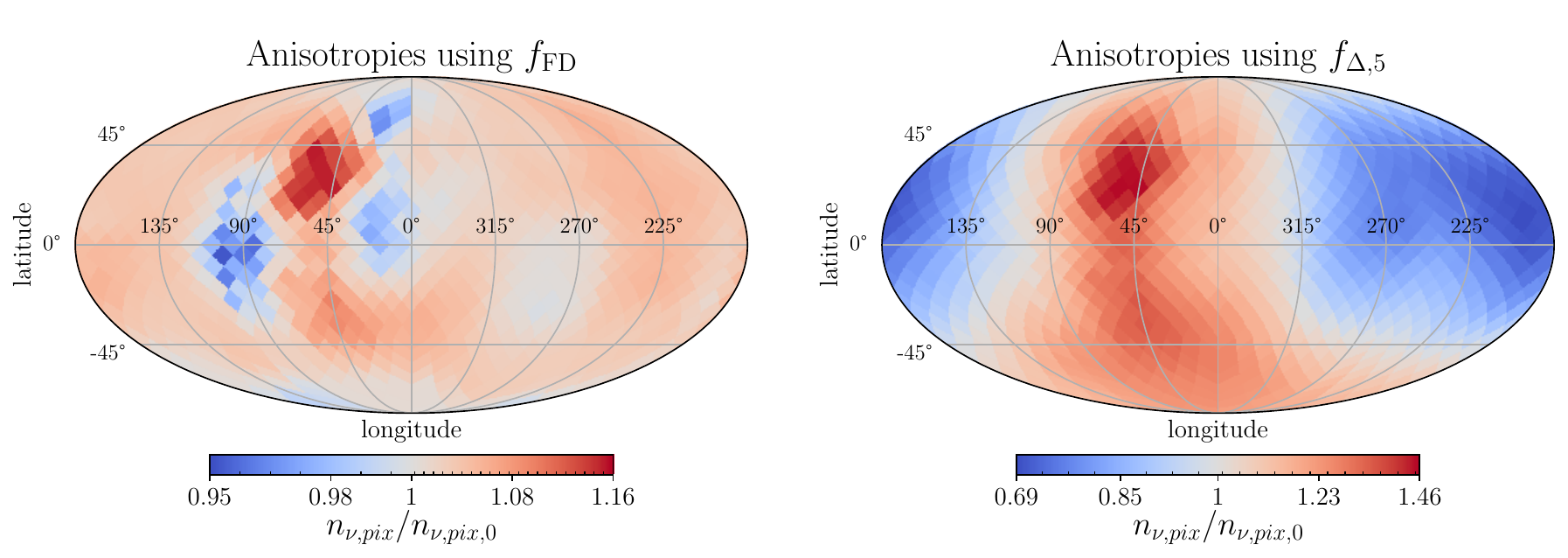}
    \caption{Comparison of C$\nu$B anisotropy sky maps for $0.1 \, {\rm eV}$ neutrinos, when applying the two different boundary conditions: using the isotropic Fermi-Dirac phase-space $f_{\mathrm{FD}}$ generates the left panel, while using the specific perturbed phase-space $f_{\Delta,5}$ generates the right panel. The densities are normalized to the cosmological average value on a per pixel basis.}
    \label{fig:FD_vs_PF_skymaps_0.1eV}
\end{figure}

The final point of interest concerns the C$\nu$B anisotropies, which are also affected by the change in boundary conditions. We first visualize the effects by showing a comparison between skymaps for $0.1 \, {\rm eV}$ neutrinos in figure~\ref{fig:FD_vs_PF_skymaps_0.1eV}, where the left map was generated using $f_{\mathrm{FD}}$ and the right map with the specific realisation $f_{\Delta,5}$ used in the right panels of figure~\ref{fig:eta_bands}. One noticeable change is the slight widening of the range of the under- and overdensity values for the pixels. Even though the minimal value decreases and larger patches depict underdense regions in the sky, the overall effect of these primordial fluctuations is still net positive for the number densities, as we saw in figure~\ref{fig:eta_bands}. To illustrate the impact for lighter neutrinos, in figure~\ref{fig:FD_vs_PF_skymaps_0.01eV} of appendix~\ref{app:figures} we show a skymap for a neutrino mass of $0.01 \, {\rm eV}$.

\begin{figure}[t!]
    \centering
    \begin{subfigure}{.48\textwidth}
        \centering
        \includegraphics[width=\linewidth]{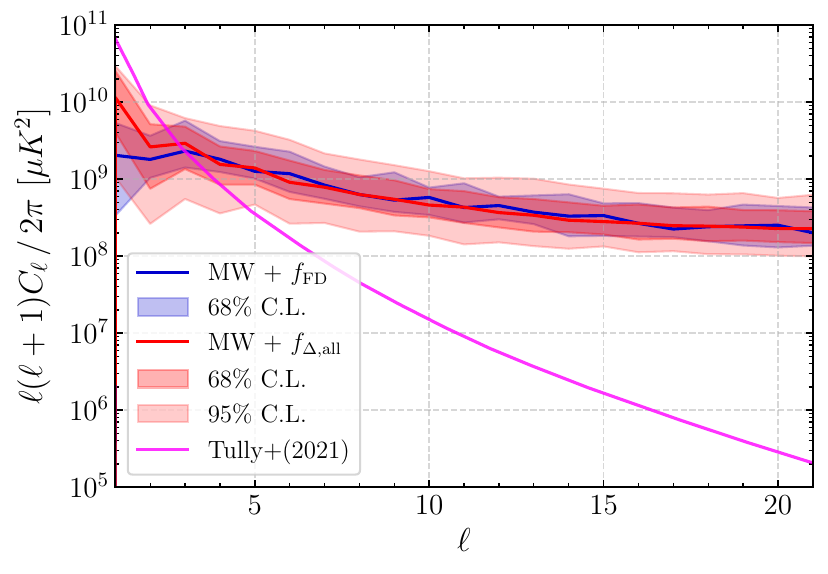}
    \end{subfigure}%
    \begin{subfigure}{.48\textwidth}
        \centering
        \includegraphics[width=\linewidth]{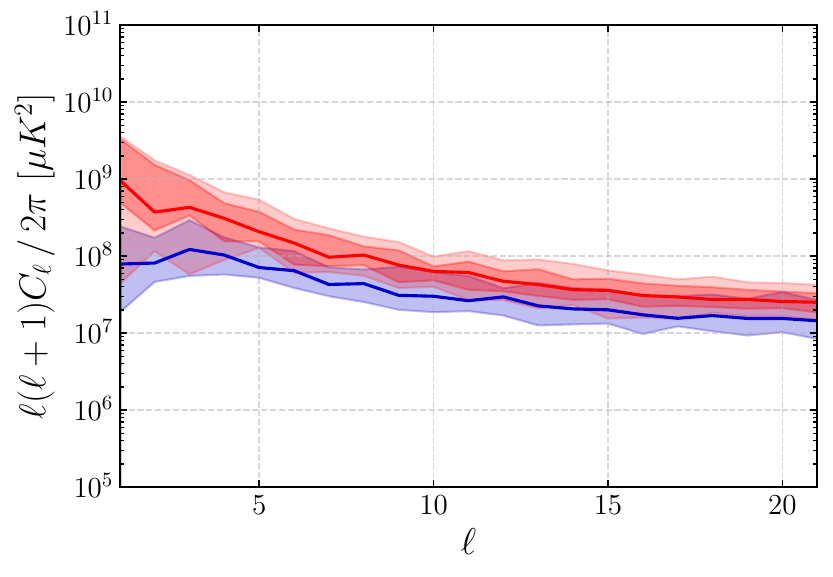}
    \end{subfigure}
    \caption{Angular power spectra of the C$\nu$B anisotropies for $0.1 \, {\rm eV}$ (left panel) and $0.01 \, {\rm eV}$ (right panel) neutrinos. The bands in blue result from using an isotropic Fermi-Dirac ($f_{\rm FD}$) as the boundary conditions, whereas all perturbed phase-spaces ($f_{\Delta,{\rm all}}$) were used to produce the red bands. The solid curves represent the median values of our DM halo sample, whereas the dark and light shaded areas indicate 68\% and 95\% coverage bands, respectively. The power spectrum for $0.1$ eV neutrinos from \cite{Tully:2021key} is shown in magenta for comparison.}
    \label{fig:power_spectra_all}
\end{figure}

Another change is the appearance of features on large angles. This  becomes more apparent in the corresponding angular power spectra, that we show in Figure~\ref{fig:power_spectra_all} for $0.1 \, {\rm eV}$ (left panel) and $0.01 \, {\rm eV}$ (right panel) neutrinos. The blue and red band in this figure denote the spectrum of the sky map generated using $f_{\mathrm{FD}}$ (with uncertainty coming from only the DM halos) and $f_{\Delta,{\rm all}}$ (representing the total uncertainties), respectively.
In figure~\ref{fig:power_spectra_split} of appendix~\ref{app:figures}, we seperately show the effects of these two sources of uncertainty, and conclude that the different realizations of $f_\Delta$ dominate the uncertainties.
For both neutrino masses, the power for the first few multipoles can increase up to an order of magnitude when including the effects of primordial fluctuations. The power for higher multipoles are only affected for the lighter neutrinos. This reflects the fact that the lighter neutrinos traverse larger distances before reaching Earth, therefore having more time to pick up the effects on smaller scales than the heavier neutrinos.\footnote{The pixel resolution is about $\theta_{\rm pix} \approx 0.016 \, {\rm sr}$ (${\sim}1 \, {\rm deg}^2$), so our skymaps are valid up to $\ell \approx \pi/\theta_{\rm pix} \approx 24$.}

In~\cite{Mertsch:2019qjv}, it was found that most of the gravitational clustering happens at very low redshifts of $z \lesssim 0.5$, and we tested that the same holds true for our framework. Moreover, we have found that lowering the simulation boundary redshift to about $z_{\rm sim} \approx 0.5$, and hence allowing more time for the primordial overdensities to grow while still simulating the most important clustering regime with our N-1-body framework, does not yield a significant increase in power at low-$\ell$. We show this explicitly in figure~\ref{fig:power_spectra_0.1_z0p5eV} of appendix~\ref{app:figures}. This is in line with the findings of~\cite{Tully:2021key} where it was shown that the greatest contribution of primordial fluctuations to the low-$\ell$ part of the C$\nu$B spectra comes from scales below ${\sim}100 \, {\rm Mpc}$ for $0.05 \, {\rm eV}$ neutrinos (and hence even smaller scales for more massive neutrinos), i.e. at very low redshifts.

\section{Discussion \& Conclusions}
\label{sec:discussion_and_conclusions}

In this last section we discuss some advantages and shortcomings of this work and additional points of interest regarding the results. We end by discussing and summarizing our main findings.

\paragraph{Pros and cons.} A shortcoming of modelling LSS with linear theory based on power spectra is that it is unconstrained. The resulting, and effectively random large scale temperature fluctuations may not exhibit the pattern realized in nature. So we can only make statements about average quantities and discuss global trends, as done in this work.
An advantage of this simulation framework is its speed and flexibility: number densities can quickly be compared for different phase-space or DM distributions. Future constraints on these two quantities may make this simulation approach more useful.

\paragraph{Comparison to state-of-the-art.} Our results differ from those of the latest N-body simulation of~\cite{Elbers:2023mdr}, especially the magnitude and shape of the anisotropy power spectra. The median power on large scales in figure~\ref{fig:power_spectra_all} is about an order of magnitude lower compared to the results of that study, whereas our power spectra drop off less quickly on smaller scales. Both are explicable by looking at how our methodology differs: the smallest cells in our simulations have a length of ${\sim}1.5 \, {\rm kpc}$\footnote{The {\sc TangoSIDM} simulation suite has a resolution floor of $650 \, {\rm pc}$, but due to our computational availability we rediscretized our simulation space, resolving scales down to ${\sim}1.5 \, {\rm kpc}$.}, but we don't model the effects of LSS in the non-linear regime for redshift $\lesssim 0.5$. The result is more power on smaller scales but less on the largest scales.

\paragraph{Importance of primordial fluctuations.} The idea of introducing small phase-space density fluctuations at the boundary of the simulation to account for any clustering that took place since the onset of structure formation has been mentioned in~\cite{Mertsch:2019qjv}. However, based on their observation that their results were independent from pushing back the redshift of their simulation boundary, the authors concluded that any clustering since the onset of structure formation outside their simulation domain could be safely neglected. As we showed in e.g. figure~\ref{fig:eta_bands}, the contribution of these seemingly miniscule fluctuations do contribute to C$\nu$B properties, and hence should not be ignored in such a simulation framework. 

\paragraph{Future detectability.} A direct detection of the C$\nu$B via experiments like PTOLEMY, and its proposed successors using polarized tritium that may be sensitive to the incoming direction of the relic neutrinos~\cite{Lisanti:2014pqa}, may be possible in the future. However, it is expected that the initially poor angular resolution of such directional C$\nu$B experiments will only be able to pick up the variations on large scales, but as these scales would be sensitive to changes in the C$\nu$B temperature, that quantity could be measurable~\cite{Tully:2021key}. Since the primordial fluctuations investigated in this work mainly affect the largest scales, they would need to be included in the modeling if an inference of the C$\nu$B temperature is to be attempted.
\\

In this work we extended our C$\nu$B simulator developed in~\cite{Zimmer:2023jbb} that only included local DM halos resembling the MW in terms of virial mass. By using cosmological perturbation theory to compute the evolution of C$\nu$B temperature anisotropies, we could generate a perturbed phase-space distribution for the neutrinos at our simulation boundary, giving more realistic initial conditions than the usual isotropic Fermi-Dirac distribution. The key points to be taken away from this work are as follows.
\begin{itemize}
    \item The redshift boundary of an N-1-body simulation framework such as the one used here matters, as the neutrino overdensities can grow in magnitude such that they have to be modelled and accounted for accurately. We show how the neutrino density constrast evolves over time in figure~\ref{fig:Delta_vs_z}.
    \item The inclusion of LSS can contribute positively to the local C$\nu$B number density for certain LSS configurations, where then linear LSS alone can account already for almost up to $20\%$ of the clustering factors for the higher neutrino masses around $0.2$ eV, as can be seen in the bottom right panel of figure~\ref{fig:eta_bands}. On the other hand, for other configurations there can be small deficits for lower neutrino masses. One way to  significantly reduce this scatter would be to use constrained realizations of the primordial matter field base on galaxy data, as done in the approach known as ``Bayesian Origin Reconstruction from Galaxies'' (BORG) \cite{Jasche:2012kq,Jasche:2018oym,Stopyra:2023yqm}, which was applied to the N-body simulations of \cite{Elbers:2023mdr}.
    \item The lowest multipoles of the C$\nu$B angular power spectra increase in power up to an order of magnitude when using the perturbed phase-space distribution $f_\Delta$ at the simulation boundary, as can be seen in figure~\ref{fig:power_spectra_all}. Since the values of $f_\Delta$ are generated in a random fashion from the $C_\ell$, we also show the spread due to different $f_\Delta$  realisations in figure~\ref{fig:power_spectra_split}, where it is visible that the increase in power can vary up to an order of magnitude.
    \item The dominant contribution to the lower multipoles of the C$\nu$B angular power spectrum comes from the structure the neutrinos traverse in their last few $\mathcal{O}(10)\text{--}\mathcal{O}(100) \, {\rm Mpc}$. As this happens at very low redshifts of $\lesssim 0.5$, lowering the redshift of our simulation boundary did not increase that power significantly, which is visible when comparing figure~\ref{fig:power_spectra_0.1_z0p5eV} to figure~\ref{fig:power_spectra_split}. 
\end{itemize}

The main motivation of this study was to find an effective way to include the effects of LSS in an N-1-body simulation framework. Figure~\ref{fig:power_spectra_all} shows that the methodology proposed here is able to partially do so: the difference in power on low multipoles can be as small as half an order of magnitude compared to the results from state-of-the-art N-body simulations~\cite{Elbers:2023mdr}, for some of our DM halos. The power towards smaller scales does not drop off as quickly, due to the nature of our highly-resolved N-1-body simulation space on MW scales. Therefore, it offers to be a viable alternative to simulate and test physical models involving the C$\nu$B in a computationally less expensive way.

\acknowledgments

The simulations for this work were performed on the Snellius Computing Clusters at SURFsara. This publication is part of the project ``One second after the Big Bang'' NWA.1292.19.231 which is financed by the Dutch Research Council (NWO). SA was partly supported by MEXT KAKENHI Grant Numbers, JP20H05850, JP20H05861, and JP24K07039. GFA is supported by the European Research Council (ERC) under the European Union's Horizon 2020 research and innovation programme (Grant agreement No. 864035 - Undark).

\appendix

\section{Supplementary figures}
\label{app:figures}

\begin{figure}[h!]
    \centering
    \begin{subfigure}{.48\textwidth}
        \centering
        \includegraphics[width=\linewidth]{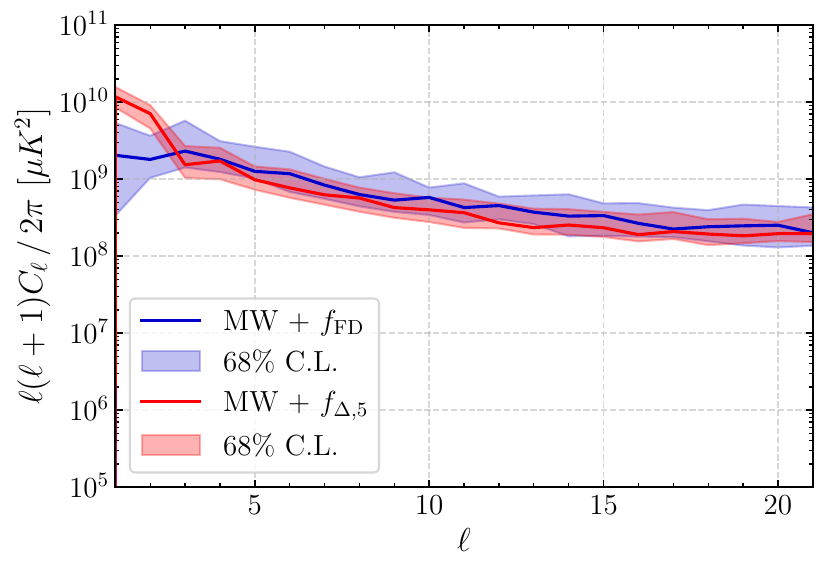}
    \end{subfigure}%
    \begin{subfigure}{.48\textwidth}
        \centering
        \includegraphics[width=\linewidth]{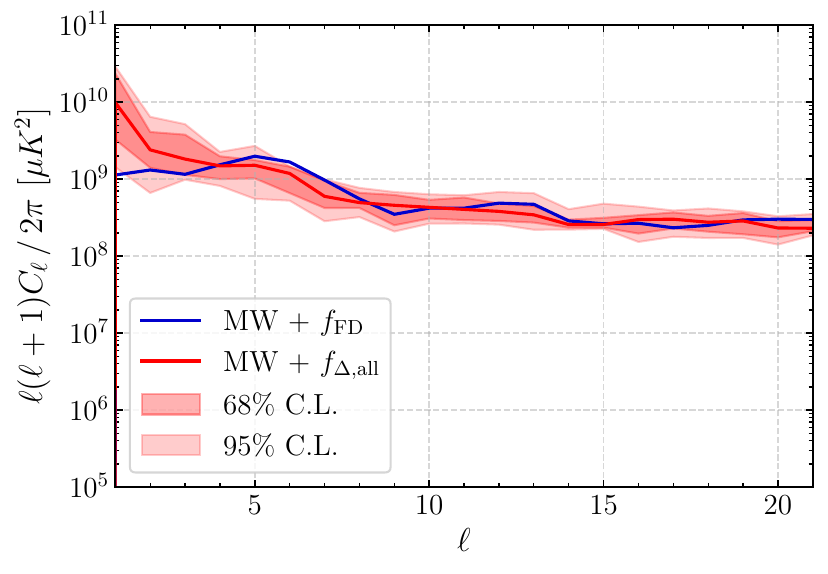}
    \end{subfigure}
    \caption{Angular power spectra of the C$\nu$B anisotropies for $0.1 \, {\rm eV}$ neutrinos, with the same color code as in  figure~\ref{fig:power_spectra_all}.  \textbf{Left panel:} The dispersion is obtained from our whole halo sample and one realization of $f_\Delta$. \textbf{Right panel:} The dispersion is obtained from 10 different realizations of $f_\Delta$ and one halo. This shows that the uncertainty is dominated by the $f_\Delta$'s, since the associated effects can vary up to an order of magnitude at low-$\ell$.}
    \label{fig:power_spectra_split}
\end{figure}

\begin{figure}[h!]
    \centering
    \begin{subfigure}{.48\textwidth}
        \centering
        \includegraphics[width=\linewidth]{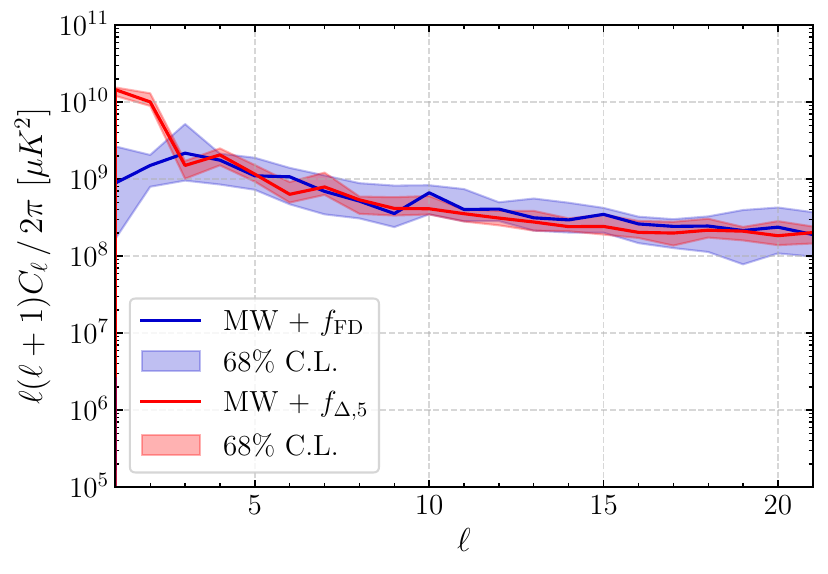}
    \end{subfigure}%
    \begin{subfigure}{.48\textwidth}
        \centering
        \includegraphics[width=\linewidth]{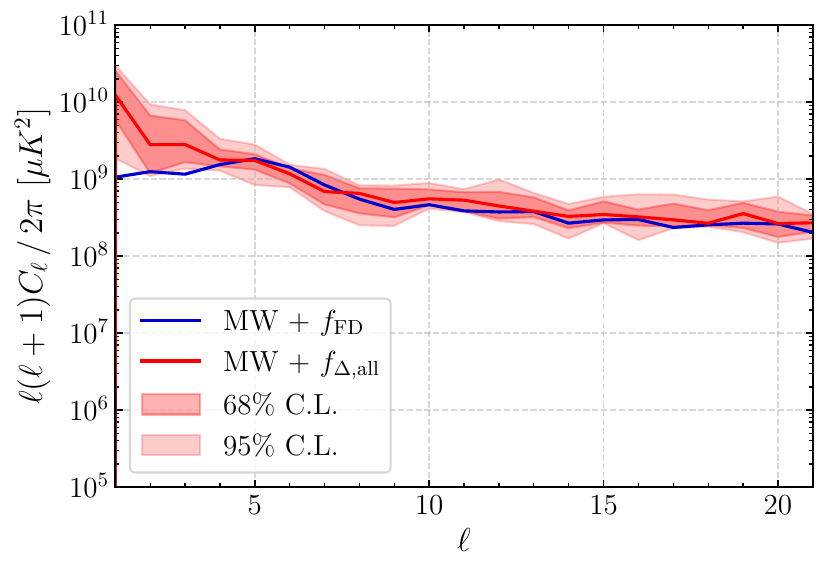}
    \end{subfigure}
    \begin{subfigure}{.48\textwidth}
        \centering
        \includegraphics[width=\linewidth]{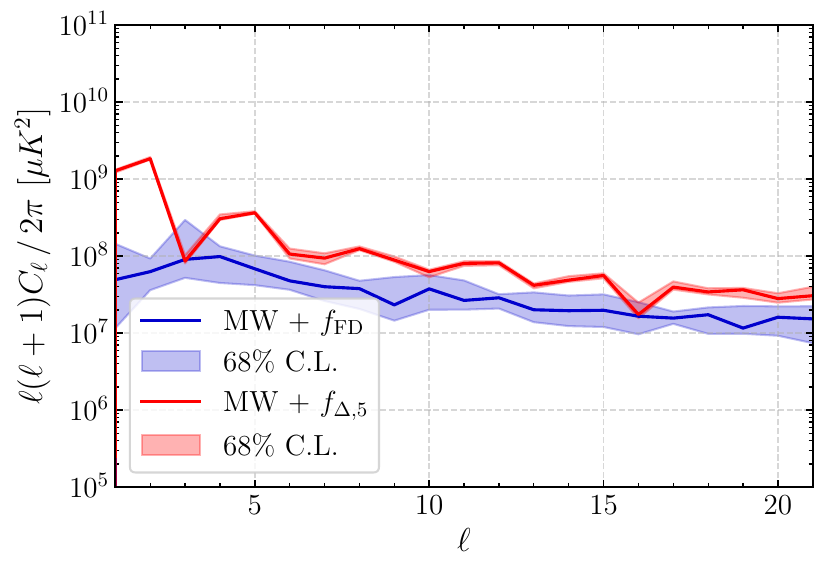}
    \end{subfigure}%
    \begin{subfigure}{.48\textwidth}
        \centering
        \includegraphics[width=\linewidth]{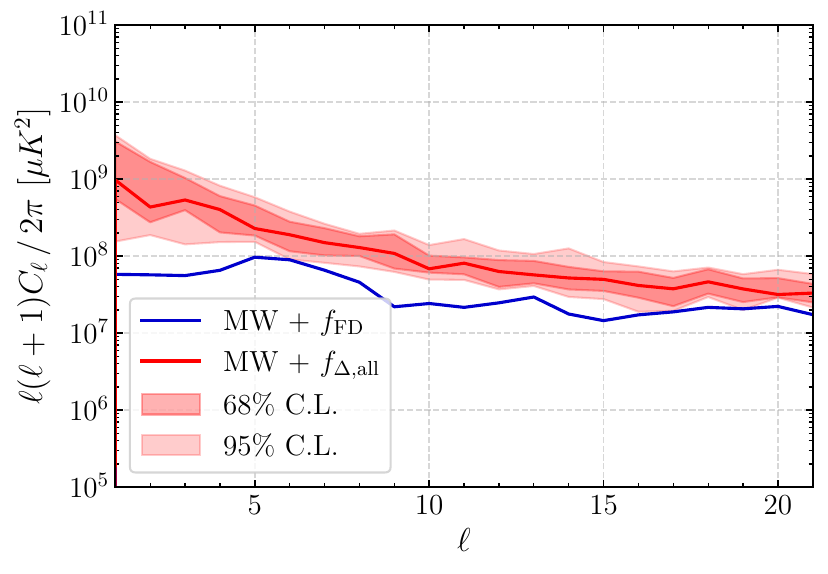}
    \end{subfigure}
    \caption{Angular power spectra of the C$\nu$B anisotropies for $0.1 \, {\rm eV}$ (top row) and $0.01 \, {\rm eV}$ (bottom row) neutrinos, when lowering the redshift of the simulation boundary from $z_{\rm sim} = 4$  to $z_{\rm sim} = 0.5$. The meaning of the uncertainty bands is the same as in figure~\ref{fig:power_spectra_split}.}
    \label{fig:power_spectra_0.1_z0p5eV}
\end{figure}

\begin{figure}[h!]
    \centering
    \includegraphics[width=\textwidth]{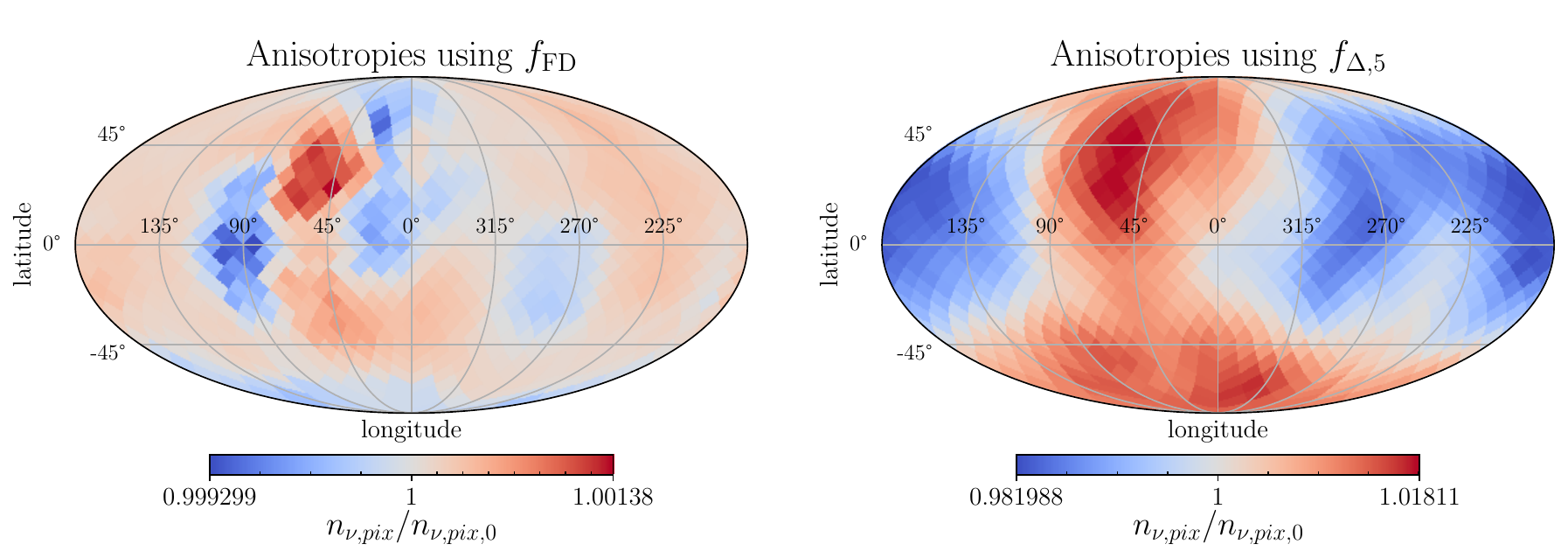}
    \caption{Comparison of C$\nu$B anisotropy sky maps as in figure~\ref{fig:FD_vs_PF_skymaps_0.1eV}, but for $0.01 \, {\rm eV}$ neutrinos.}
    \label{fig:FD_vs_PF_skymaps_0.01eV}
\end{figure}


\providecommand{\href}[2]{#2}\begingroup\raggedright\endgroup

\end{document}